\documentclass[11pt,english]{article}
\usepackage{graphicx}
\usepackage[T1]{fontenc}
\usepackage[latin9]{inputenc}
\usepackage[letterpaper]{geometry}
\usepackage{float}
\usepackage{amsbsy}
\usepackage{setspace}
\usepackage{amssymb}
\usepackage{esint}
\makeatletter

\floatstyle{ruled}
\newfloat{algorithm}{tbp}{loa}
\floatname{algorithm}{Algorithm}

\usepackage{babel}

\begin{document}

\title{Spin Glasses and Nonlinear Constraints in Portfolio Optimization}

\author{M. Andrecut}

\maketitle
{

\centering Unlimited Analytics Inc.

\centering Calgary, AB, Canada

\centering mircea.andrecut@gmail.com

} 
\begin{abstract}
We discuss the portfolio optimization problem with the obligatory
deposits constraint. Recently it has been shown that as a consequence of this
nonlinear constraint, the solution consists of an exponentially large
number of optimal portfolios, completely different from each other,
and extremely sensitive to any changes in the input parameters of
the problem, making the concept of rational decision making questionable.
Here we reformulate the problem using a quadratic obligatory
deposits constraint, and we show that from the physics point
of view, finding an optimal portfolio amounts to calculating the mean-field
magnetizations of a random Ising model with the constraint of a constant
magnetization norm. We show that the model reduces to an eigenproblem,
with $2N$ solutions, where $N$ is the number of assets defining
the portfolio. Also, in order to illustrate our results, we present 
a detailed numerical example of a portfolio of several risky common
stocks traded on the Nasdaq Market.  
\end{abstract}

\section{Introduction}

Portfolio optimization is an important problem in economic analysis
and risk management \cite{key-1,key-2}, and under certain nonlinear constraints
maps exactly into the problem of finding the ground states of a long-range
spin glass \cite{key-3,key-4,key-5}. The main assumption is that the return of any
financial asset is described by a random variable, whose expected
mean and variance are interpreted as the reward, and respectively
the risk of the investment. The problem can be formulated as following:
given a set of financial assets, characterized by their expected mean
and their covariances, find the optimal weight of each asset, such
that the overall portfolio provides the smallest risk for a given
overall return. The standard mean-variance optimization problem has
an unique solution describing the so called ``efficient frontier''
in the $(risk,return)$-plane \cite{key-6}. The expected return is a monotonically
increasing function of the standard deviation (risk), and for accepting
a larger risk the investor is rewarded with a higher expected return.
Recently it has been shown that the portfolio optimization problem
containing short sales with obligatory deposits (margin accounts)
is equivalent to the problem of finding the ground states of a long-range
Ising spin glass, where the coupling constants are related to the
covariance matrix of the assets defining the portfolio \cite{key-3,key-4,key-5}.
As a consequence of this nonlinear constraint, the solution consists
of an exponentially large number of optimal portfolios, completely
different from each other, and extremely sensitive to any changes
in the input parameters of the problem. Therefore, under such constraints,
the concept of rational decision making becomes questionable, since
the investor has an exponential number of ``options'' to choose
from. Here, we discuss the portfolio optimization problem using a
quadratic formulation of the nonlinear obligatory deposits constraint.
From the physics point of view, finding an optimal portfolio amounts
to calculating the mean-field magnetizations of a random Ising model
with the constraint of a constant magnetization norm. We show that
the proposed model reduces to an eigenproblem, with $2N$ solutions,
where $N$ is the number of assets defining the portfolio. In support
to our results, we also work out a detailed numerical example of a
portfolio of several risky common stocks traded on the Nasdaq Market.

\section{Nonlinear optimization model}

A portfolio is an investment made in $N$ assets $A_{n}$, with the
expected returns $r_{n}$, and covariances $s_{nm}=s_{mn}$, $n,m=1,2,...,N$.
Let $w_{n}$ denote the relative amount invested in the $n$-th asset.
Negative values of $w_{n}$ can be interpreted as short selling. The
variance of the portfolio captures the risk of the investment, and
it is given by:
\begin{equation}
s^{2}=\sum_{i=1}^{N}\sum_{j=1}^{N}w_{i}w_{j}s_{ij}=\mathbf{w}^{T}\mathbf{S}\mathbf{w},
\end{equation}
where $\mathbf{w}=[w_{1},w_{2},\ldots,w_{N}]^{T}$ is the vector of weights,
and $\mathbf{S}=[s_{nm}]$ is the covariance matrix. Also, another
characteristic of the portfolio is the expected return:
\begin{equation}
\rho=\sum_{n=1}^{N}w_{n}r_{n}=\mathbf{w}^{T}\mathbf{r},
\end{equation}
where $\mathbf{r}=[r_{1},r_{2},\ldots,r_{N}]^{T}$ is the vector of asset returns. The standard portfolio
selection problem consists in finding the solution of the following
multi-objective optimization problem {[}1,2,6{]}:
\begin{equation}
\min_{\mathbf{w}}\left\{ s^{2}=\mathbf{w}^{T}\mathbf{S}\mathbf{w}\right\} ,
\end{equation}
\begin{equation}
\max_{\mathbf{w}}\left\{ \rho=\mathbf{w}^{T}\mathbf{r}\right\} ,
\end{equation}
subject to the invested wealth constraint: 
\begin{equation}
\sum_{n=1}^{N}w_{n}=1.
\end{equation}
As mentioned in the introduction, this problem has an unique solution,
which can be obtained using the method of Lagrange multipliers \cite{key-1,key-2,key-6}. 

Recently it has been shown that by replacing the invested wealth constraint
(5) with an obligatory deposits constraint the problem cannot be solved
analytically anymore {[}3-5{]}. The constraint consists in imposing
the requirement to leave a certain deposit (margin) proportional to
the value of the underlying asset, and it has the form:
\begin{equation}
\gamma\sum_{n=1}^{N}\left|w_{n}\right|=W,
\end{equation}
where $\gamma>0$ is the fraction defining the margin requirement,
and $W$ is the total wealth invested. As a direct consequence of
the constraint's nonlinearity, the problem has an exponentially large
number of solutions:
\begin{equation}
n(N,\rho)\sim\exp(\omega(\rho)N),
\end{equation}
where $\omega(\rho)$ is a positive number depending on the portfolio
return \cite{key-3,key-4,key-5}. The solutions are also completely different from
each other, and extremely sensitive to any changes in the input parameters
of the problem. Thus, finding the global optimum becomes prohibitive
(NP-problem) for a larger $N$. 

Let us now to reformulate this constraint using a quadratic function:
\begin{equation}
\gamma\sum_{n=1}^{N}w_{n}^{2}=W.
\end{equation}
Thus, we impose the requirement to leave a certain deposit proportional
to the quadratic value of the asset. This is equivalent to a constant
norm $\left\Vert \mathbf{w}\right\Vert ^{2}=k=W/\gamma$. Also, we
combine the multi-objective optimization problem into a single Lagrangian
objective function as following:
\begin{equation}
\min_{\mathbf{w},\mu}\left\{ F(\mathbf{w},\lambda,\mu)=\lambda\mathbf{w}^{T}\mathbf{S}\mathbf{w}-(1-\lambda)\mathbf{w}^{T}\mathbf{r}-\mu(\mathbf{w}^{T}\mathbf{w}-k)\right\} ,
\end{equation}
where $\lambda\in[0,1]$ is the risk aversion parameter, and $\mu$
is the Lagrange parameter. 

If $\lambda=0$ then the solution corresponds to the portfolio with
maximum return, without considering the risk. In this case the optimal
solution will be formed only by the asset with the greatest expected
return. The case with $\lambda=1$ corresponds to the portfolio with
minimum risk, regardless the value of the expected return. In this
case the problem becomes: 
\begin{equation}
\min_{\mathbf{w},\mu}\left\{ F(\mathbf{w},1,\mu)=\mathbf{w}^{T}\mathbf{S}\mathbf{w}-\mu(\mathbf{w}^{T}\mathbf{w}-k)\right\} ,
\end{equation}
with the solutions given by the equation:
\begin{equation}
\nabla_{\mathbf{w}}F(\mathbf{w},\lambda,\mu)=2\mathbf{S}\mathbf{w}-2\mu\mathbf{w}=0.
\end{equation}
This is a standard eigenproblem:
\begin{equation}
\mathbf{S}\mathbf{w}=\mu\mathbf{w},
\end{equation}
where $\mathbf{S}$ is a symmetric matrix with $N$ real eigenvalues,
and $N$ real eigenvectors. The eigenvector corresponding to the largest
eigenvalue will provide the global optimum, since it will have the
lowest risk. 

Any value $\lambda\in(0,1)$ represents a tradeoff between the risk
and return. In this case the solution corresponds to the critical
point of the Lagrangian, which is also the solution of the following
system of equations:
\begin{equation}
\nabla_{\mathbf{w}}F(\mathbf{w},\lambda,\mu)=2\lambda\mathbf{S}\mathbf{w}-(1-\lambda)\mathbf{r}-2\mu\mathbf{w}=0,
\end{equation}
\begin{equation}
\frac{\partial F(\mathbf{w},\lambda,\mu)}{\partial\mu}=\mathbf{w}^{T}\mathbf{w}-k=0.
\end{equation}
One can see that the Lagrangian objective function is equivalent to
the free energy of an Ising model with random couplings $J_{nm}=-2\lambda s_{nm}$
and a random magnetic field $h_{n}=(1-\lambda)r_{n}$. From the physics
point of view, finding an optimal portfolio amounts to calculating
the mean-field magnetizations $w_{n}$ of this random Ising model
with the constraint of a constant magnetization norm. In the following
we show that solving this system of equations reduces to an inhomogeneous
eigenproblem. 

From the first equation we have:
\begin{equation}
\mathbf{w}=\frac{1}{2}(1-\lambda)(\lambda\mathbf{S}-\mu\mathbf{I})^{-1}\mathbf{r}.
\end{equation}
Introducing this result into the second equation we obtain:
\begin{equation}
\frac{1}{4}(1-\lambda)^{2}\mathbf{r}^{T}(\lambda\mathbf{S}-\mu\mathbf{I})^{-2}\mathbf{r}-1=0.
\end{equation}
The left-hand side of this equation is the Schur complement of the
matrix:
\begin{equation}
\mathbf{M}=\left[\begin{array}{cc}
(\lambda\mathbf{S}-\mu\mathbf{I})^{2} & \frac{1}{2}(1-\lambda)\mathbf{r}\\
\frac{1}{2}(1-\lambda)\mathbf{r}^{T} & 1
\end{array}\right].
\end{equation}
Since this matrix must be singular (the Schur complement is zero),
we have:
\begin{equation}
\mathrm{det}\left[(\lambda\mathbf{S}-\mu\mathbf{I})^{2}-\frac{1}{4}(1-\lambda)^{2}\mathbf{r}\mathbf{r}^{T}\right]=0,
\end{equation}
which reduces to:
\begin{equation}
\mathrm{det}\left[\frac{1}{4}(1-\lambda)^{2}\mathbf{r}\mathbf{r}^{T}-\lambda^{2}\mathbf{S}^{2}+2\lambda\mu\mathbf{S}-\mu^{2}\mathbf{I}\right]=0.
\end{equation}
Obviously, there is a vector $\mathbf{w}$ such that:
\begin{equation}
\left[\frac{1}{4}(1-\lambda)^{2}\mathbf{r}\mathbf{r}^{T}-\lambda^{2}\mathbf{S}^{2}+2\lambda\mu\mathbf{S}-\mu^{2}\mathbf{I}\right]\mathbf{w}=0.
\end{equation}
This is an inhomogeneous $N\times N$ eigenproblem \cite{key-7}, and it can be
reduced further to a $2N\times2N$ standard eigenproblem by introducing
the following quantity:
\begin{equation}
\mathbf{u}=\mu\mathbf{w},
\end{equation}
such that we have:
\begin{equation}
\left[\frac{1}{4}(1-\lambda)^{2}\mathbf{r}\mathbf{r}^{T}-\lambda^{2}\mathbf{S}^{2}\right]\mathbf{w}+2\lambda\mathbf{S}\mathbf{u=\mu}\mathbf{u}.
\end{equation}
By combining the last two equations into a matrix representation we
obtain:
\begin{equation}
\mathbf{\left[\begin{array}{cc}
\mathbf{0} & \mathbf{I}\\
\mathbf{A} & \mathbf{B}
\end{array}\right]}\left[\begin{array}{c}
\mathbf{w}\\
\mathbf{u}
\end{array}\right]=\mu\left[\begin{array}{c}
\mathbf{w}\\
\mathbf{u}
\end{array}\right],
\end{equation}
where
\begin{equation}
\mathbf{A}=\frac{1}{4}(1-\lambda)^{2}\mathbf{r}\mathbf{r}^{T}-\lambda^{2}\mathbf{S}^{2},
\end{equation}
and
\begin{equation}
\mathbf{B}=2\lambda\mathbf{S},
\end{equation}
and $\mathbf{0}$, $\mathbf{I}$ are the zero, and respectively identity
matrices. This eigenproblem obviously has $2N$ eigenvalues $\mu$,
that may be real or complex. 

If $\mathbf{w}$ is a real optimal portfolio, associated to a real
eigenvalue, then the corresponding risk and return are given by:
\begin{equation}
s=\sqrt{\mathbf{w}^{T}\mathbf{S}\mathbf{w}},\quad\rho=\left|\mathbf{w}^{T}\mathbf{r}\right|.
\end{equation}
In the above equation we have defined the return as the absolute value
of the scalar product. This is a consequence of the fact that the
eigenvectors can be determined only up to the sign value, i.e. both
$[\mathbf{w},\mathbf{u}]^{T}$ and $-[\mathbf{w},\mathbf{u}]^{T}$
are eigen-vectors corresponding to the same eigenvalue $\mu$. This
means that if the return is negative, we can simply change the sign
$(\mathbf{w}\rightarrow-\mathbf{w})$, such that the return becomes
positive. 

The real and imaginary parts of the complex portfolios are also valid
investment portfolios, however they are sub-optimal and may be discarded.
Assuming that $\mathbf{w}=\mathbf{x}+i\mathbf{y}$, where $\mathbf{x}=\mathrm{Re}(\mathbf{w})$
and $\mathbf{y}=\mathrm{Im}(\mathbf{w})$, the risk and return of
the real and respectively imaginary parts can be determined as following:
\begin{equation}
s_{x}=\sqrt{\mathbf{x}^{T}\mathbf{S}\mathbf{x}},\quad\rho_{x}=\left|\mathbf{x}^{T}\mathbf{r}\right|,
\end{equation}
\begin{equation}
s_{y}=\sqrt{\mathbf{y}^{T}\mathbf{S}\mathbf{y}},\quad\rho_{y}=\left|\mathbf{y}^{T}\mathbf{r}\right|.
\end{equation}

\begin{figure}[ht]
\centering \includegraphics[scale=0.7]{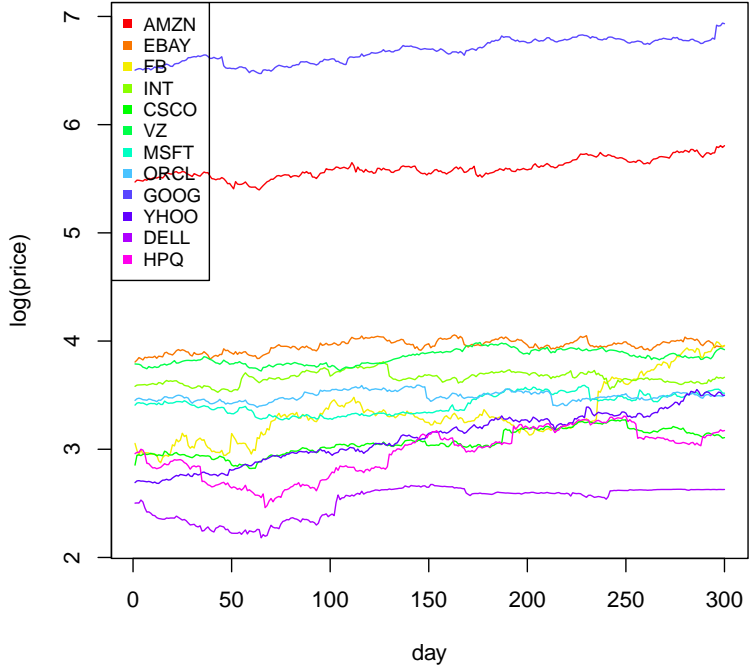}\caption{Asset prices (log scale).}
\end{figure}
\begin{figure}[ht]
\centering \includegraphics[scale=0.7]{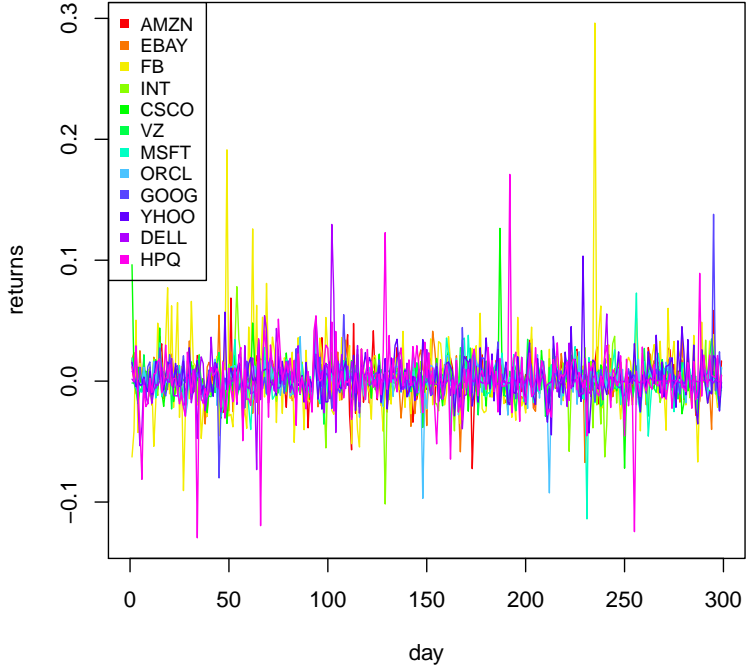}
\caption{Daily returns of the assets.}
\end{figure}

\section{Numerical example}

In order to illustrate the above results, we consider the case of
a portfolio consisting of $N=12$ common stocks from IT industry:
\begin{equation}
\mathrm{AMZN,EBAY,FB,INT,CSCO,VZ,MSFT,ORCL,GOOG,YHOO,DELL,HPQ}
\end{equation}
A historical record of daily prices of these stocks for the last $T=300$
trading days was used to estimate the mean return and the covariance
matrix. 

The daily returns of the assets are calculated as: 
\begin{equation}
R(n,t)=\frac{P(n,t+1)-P(n,t)}{P(n,t)},
\end{equation}
where $t=1,2,...,T$ is the day index, and $P(n,t)$ is the price
of asset $A_{n}$ at the closing day $t$. The estimate average returns
and covariances are:
\begin{equation}
r_{n}=\frac{1}{T}\sum_{t=1}^{T}R(n,t),\quad n=1,2,...,N,
\end{equation}
\begin{equation}
s_{ij}=\frac{1}{T}\sum_{t=1}^{T}[R(i,t)-r_{i}][R(j,t)-r_{j}],\quad i,j=1,2,...,N.
\end{equation}
The time series of the asset prices for the considered time period
are given in Figure 1. In Figure 2 we give also the expected daily
returns of the assets. 

The risk aversion parameter $\lambda$ was discretized as $\lambda_{t}=t/T$,
$t=0,1,...,T$, where $T=1000$. Also, the fraction defining the margin requirement 
was set to $\gamma=1$. Thus by solving the eigenproblem for each value of $\lambda$, 
one obtains $2NT=24,000$ eigenvalues
and eigenvectors containing the weights of the portfolios. The risk-return,
$(s,\rho)$, representation of these complex solutions is shown in
Figure 3, which also shows the risk-return values of the stocks included
in the portfolio. The real contributions $(s_{x},\rho_{x})$ are shown
in blue, while the imaginary contributions $(s_{y},\rho_{y})$ are shown
in red. The pure real solutions, corresponding to the real eigenvalues
are extracted in Figure 4. Here we also show the portfolio with the
maximum Sharpe ratio $\xi=\rho/s$, The Sharpe ratio represents the
expected return per unit of risk [1,2]. The portfolio with maximum Sharpe
ratio $\xi$ gives the highest expected return per unit of risk, and
therefore is the most \textquotedbl{}risk-efficient\textquotedbl{}
portfolio.

\section{Conclusion}

We have considered the portfolio optimization problem with the obligatory
deposits constraint, when both long buying and short selling 
of a relatively large number of assets
is allowed. Recently it has been shown that as a consequence of this
nonlinear constraint, the solution consists of an exponentially large
number of optimal portfolios, completely different from each other,
and extremely sensitive to any changes in the input parameters of
the problem, making the concept of rational decision making questionable.
Here we have reformulated the problem using a quadratic obligatory
deposits constraint, and we have shown that from the physics point
of view, finding an optimal portfolio amounts to calculating the mean-field
magnetizations of a random Ising model with the constraint of a constant
magnetization norm. We have shown that the model reduces to an eigenproblem,
with $2N$ solutions, where $N$ is the number of assets defining
the portfolio. Also, in order to illustrate our results, we have presented
a detailed numerical example of a portfolio of several risky common
stocks traded on the Nasdaq Market.

\begin{figure}[ht]
\centering \includegraphics[scale=0.7]{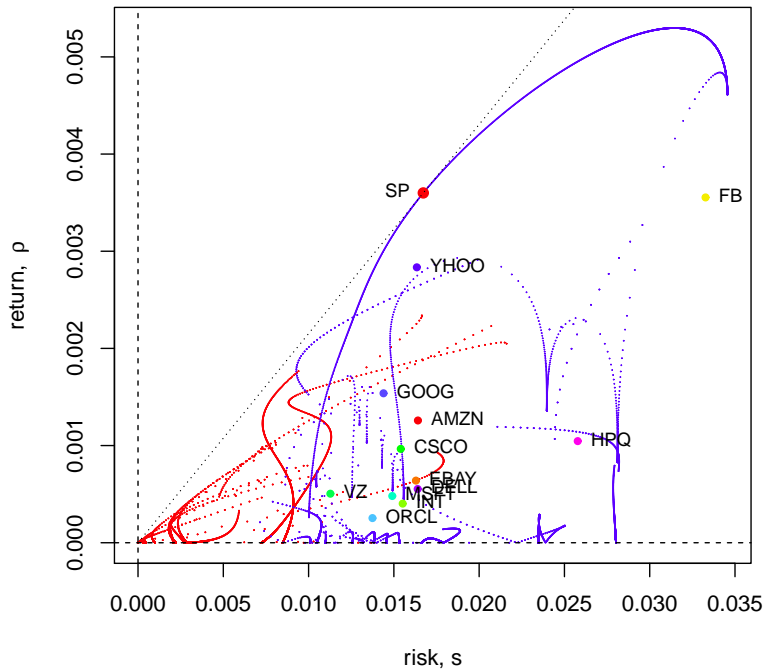}
\caption{Complex solutions of the portfolio optimization problem with quadratic obligatory deposits constraint (blue=real and red=imaginary components). SP is the portfolio with the maximum Sharpe ratio.}
\end{figure}
\begin{figure}[ht5]
\centering \includegraphics[scale=0.7]{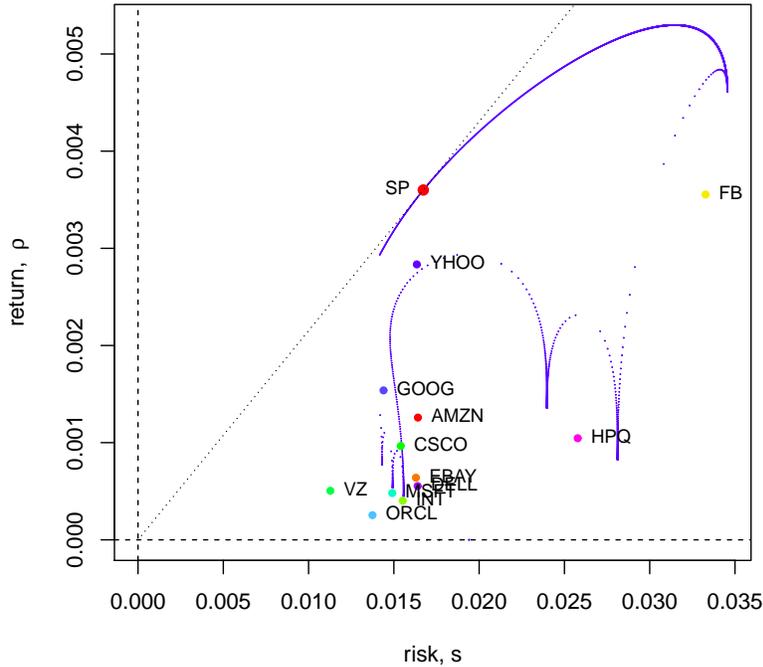}
\caption{Real solutions of the portfolio optimization problem with quadratic obligatory deposits constraint. SP is the portfolio with the maximum Sharpe ratio.}
\end{figure}


\begin{thebibliography}{9}

\bibitem{key-1}H. Markovitz, Portfolio selection: Efficient Diversification
of Investments, Wiley, New York, 1959.

\bibitem{key-2}E.J. Elton, M.J. Gruber, S.J. Brown, W.N. Goetzmann,
Modern Portfolio Theory and Investment Analysis, 8th ed., Wiley, New
York, 2010.

\bibitem{key-3}S. Galluccio, J.-P. Bouchaud, M. Potters, Rational
decisions, random matrices and spin glasses, Physica A 259, 449-456
(1998).

\bibitem{key-4}A. Gabor, I. Kondor, Portfolios with nonlinear constraints
and spin glasses, Physica A 274, 222-228 (1999).

\bibitem{key-5}L. Bongini, M. Degli Esposti, C. Giardin, A. Schianchi,
Portfolio optimization with short-selling and spin-glass, European Physics 
Journal B 27, 263\textendash{}272 (2002).

\bibitem{key-6}G. Cornuejols, R. Tutuncu, Optimization Methods in
Finance, Cambridge University Press, Cambridge, 2007. 

\bibitem{key-7}R.M.M. Mattheij, G. Soderlind, On inhomogeneous eigenvalue problems. I, 
Linear Algebra and its Applications 88-89, 507-531 (1987).

\end{thebibliography}
\end{document}